\documentclass[prb,twocolumn]{revtex4-1}

\usepackage{amssymb}
\usepackage{amsmath}
\usepackage[dvips]{graphicx}

\begin{document}

\title{How do magnetic microwires interact magnetostatically?}
\author{A. Pereira, J. C. Denardin, J. Escrig}
\affiliation{Departamento de F\'{\i}sica, Universidad de Santiago de Chile, USACH, Av.
Ecuador 3493, Santiago, Chile.}

\begin{abstract}
The magnetostatic interaction between two ferromagnetic microwires is
calculated as a function of their geometric parameters and compared with
those measured through magnetic hysteresis loops of glass-coated amorphous Fe%
$_{77.5}$Si$_{7.5}$B$_{15}$ microwires. The hysteresis loops are
characterized by well-defined Barkhausen jumps corresponding each to the
magnetization reversal of individual microwires, separated by horizontal
plateaux. It is shown that the magnetostatic interaction between them is
responsible for the appearance of these plateaux. Finally, using the
expression for the magnetostatic interaction is trivial to obtain the
interacting force between microwires. Our results are intended to provide
guidelines for the use of these microwires with technological purpose such
as the fabrication of magnetic sensors.
\end{abstract}

\maketitle

\section{Introduction}

In the last decades, soft magnetic materials have been deeply investigated.
Besides the basis scientific interest in their magnetic properties, there
are a great deal of technologic interest due to their use in sensing
applications, particularly in the fields of automotive, mobile
communication, medical and home appliance industries. \cite{Vazquez07,
CTD+04, MH07, KMP03, EAA+09} Moreover, these materials are very promising
for spintronic devices in magnetic recording media. \cite{YKN+07} \ Two
types of soft magnetic microwire families are currently studied: in-water
quenched amorphous wires with diameters of around 120 $\mu $m, \cite{OU95}
and quenched and drawn microwires with diameters ranging from around 2 to 20 
$\mu $m, \cite{ZGV+03} covered by a protective insulating glassy coat.
Bi-stable microwires are characterized by square-shaped hysteresis loops
defined by the abrupt reversal of the magnetization between two stable
remanent states. \cite{VGV05}

Although an array composed of a few ferromagnetic wires could in principle
seem a quite simple problem to study and model, it is striking to notice how
complex this problem can turn out to be. Effects of interparticle
interactions are in general complicated by the fact that the dipolar fields
depend upon the magnetization state of each element, which in turn depends
upon the fields due to adjacent elements. Therefore, the modelling of
interacting arrays of wires is often subject to strong simplifications like,
for example, modelling the wire using a one-dimensional modified classical
Ising model. \cite{SSC+00, KSS+02} Zhan \textit{et al. }\cite{ZGL+05} used
the dipole approximation including additionally a length correction. J. Vel%
\'{a}zquez and M. V\'{a}zquez \cite{VV02a, VV02b} considered each microwire
as a dipole, in a way that the axial field generated by a microwire is
proportional to its magnetization. Nevertheless, this model is merely
phenomenological since the comparison of experimental results with a
strictly dipolar model shows that the interaction in the actual case is more
intense. They have also calculated the magnetostatic field and expanded it
in multipolar terms, \cite{VPV03} showing that the non-dipolar contributions
of the field are non negligible for distances considered in experiments.
Besides, the magnetostatic interaction energy between two magnetic elements
of arbitrary shape was derived within the framework of a Fourier space
approach by Beleggia \textit{et al}. \cite{BTZ+04, BD05} Recently, a
detailed study of the magnetostatic interaction between two parallel wires
placed side by side has been shown. \cite{LEL+07, PLK+07} In spite of the
extended study of the dipolar interactions, the effect of magnetostatic
interactions on the hysteresis loops, due to the vertical displacement of
the wires, has not been studied yet.

\begin{figure}[tph]
\begin{center}
\includegraphics[width=8cm]{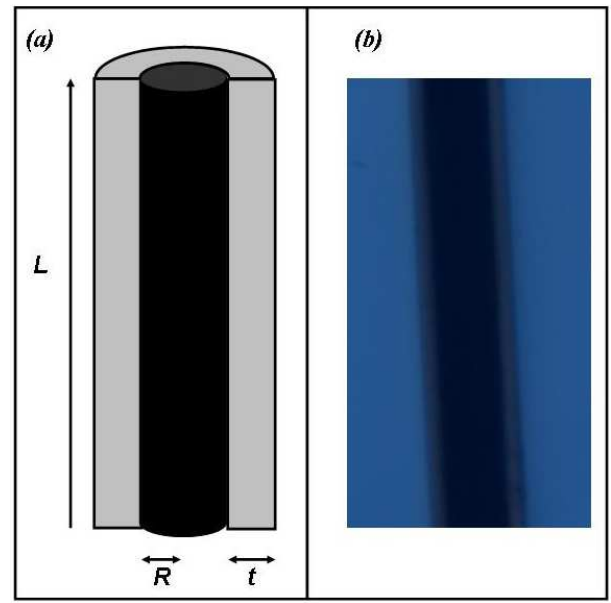}
\end{center}
\caption{(a) Geometric parameters used for the individual wire description.
(b) Micrograph image that shows a glass coated amorphous microwire.}
\end{figure}

The purpose of this work is to develop an analytical model for the full
long-range magnetostatic interaction between two microwires exploring the
possibility of varying the magnetic coupling as a function of the wires
position. The geometry of the wires is characterized by their external radii 
$R$ and length $L$, as depicted in Fig. 1. The separation between the wires
is written in terms of the interaxial distance, $d$, and the horizontal
separation, $s$, as depicted in Fig. 2. Our model goes beyond the
dipole-dipole approximation and lead us to obtain an analytical expression
for the interaction in which the lengths and radii of the wires are taken
into account. We focus on the interacting force in pairs of interacting
wires, as a function of the distance between them, in order to gain insight
on the understanding of the role of interactions on the plateaux that
appears in the hysteresis loops.

\begin{figure}[tph]
\begin{center}
\includegraphics[width=8cm]{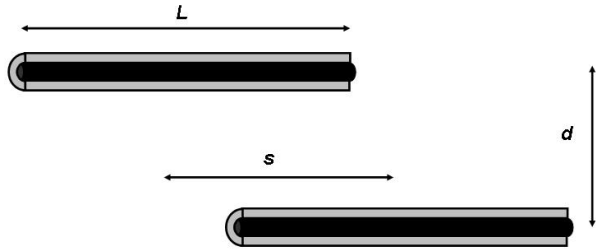}
\end{center}
\caption{Relative position of interacting wires: $d$ is the interaxial
distance and $s$ is the horizontal separation.}
\end{figure}

\section{Experimental Methods}

The experimental meaurements have been performed in glass coated amorphous
bi-stable magnetic microwires with nominal composition Fe$_{77.5}$Si$_{7.5}$B%
$_{15}$, with a saturation magnetization $M_{0}=1.2\times 10^{6}$ A/m, radii
of $R=6.5$ $\mu $m, and the thickness of the glass coating of $t=3.3$ $\mu $%
m. They are fabricated by means of Taylor-Ulitovsky technique by which the
molten metallic alloy and its glassy coating are rapidly quenched and drawn
to a kind of composite microwire. Two samples with length of $L=7$ mm each
were cut from a larger wire and fixed parallel by means of vacuum grease on
a glass holder, with a separaton of $d=45$ $\mu $m between the wires.

The morphology of microwires was investigated by a retro-optical microscope
(Olimpus). The hysteresis curves were measured on a specially designed
vibrating sample magnetometer (VSM), inserted within a pair of Helmholtz
coils. These coils have sufficient field to saturate the wires and present
the advantage of field homogeneity and the absence of remanent fields. We
will focus our attention on measurements performed at room temperature
because a low temperature there is a change in the domain structure of the
microwires, probably owing to the increasing internal stresses induced by
the different thermal expansion coefficients of the ferromagnetic alloy and
the covering glass. \cite{SSC+00}

\section{Theoretical calculations}

We adopt a simplified approach in which the discrete distribution of
magnetic moments of microwire is replaced by a continuous one characterized
by a slowly varying magnetization $\mathbf{M}(\mathbf{r})$. We consider
wires for which $L\gg R$ so it is reasonable to assume an axial
magnetization due to shape anisotropy, defined by $\mathbf{M}(\mathbf{r}%
)=M_{0}\mathbf{\hat{z}}$, where $\mathbf{\hat{z}}$ is the unit vector
parallel to the wire axis. The magnetostatic interaction between wires can
be calculated from \cite{Aharoni96}%
\begin{equation*}
E_{int}=\mu _{0}\int \mathbf{M}_{2}(\mathbf{r})\nabla U_{1}(\mathbf{r})\,dV\ 
\text{,}
\end{equation*}%
where $\mathbf{M}_{2}(\mathbf{r})$ is the magnetization of the wire $2$ and $%
U_{1}(\mathbf{r})$ is the magnetostatic potential of the microwire $1$. The
expression for this potential is given by%
\begin{equation}
U=\frac{1}{4\pi }\left( -\int_{V}\frac{\mathbf{\nabla }\cdot \mathbf{M}%
\left( \mathbf{r}^{\prime }\right) }{\left\vert \mathbf{r}-\mathbf{r}%
^{\prime }\right\vert }d^{3}r^{\prime }+\int_{S}\frac{\mathbf{\hat{n}}%
^{\prime }\cdot \mathbf{M}\left( \mathbf{r}^{\prime }\right) }{\left\vert 
\mathbf{r}-\mathbf{r}^{\prime }\right\vert }ds^{\prime }\right) \text{ .}
\end{equation}%
Note that the first term on the left-hand side of Eq. (1) vanishes, because
the magnetization field is constant; furthermore, in the surface integral of
Eq. (1) the only contributions arise from the upper and lower bases of the
wire: the upper circle is located at $z=L/2$ and the lower on at $z=-L/2$.
Due to the symmetry of the problem, we used the adecuate type of the
cylindrical kernel for the integral \cite{Jackson62}, and after few
manipulations, one finds that the integral expression for the scalar
potential is given by 
\begin{multline}
U\left( r,z\right) =\frac{M_{0}R}{2}\int_{0}^{\infty }\frac{dk}{k}%
J_{0}\left( kr\right)  \\
J_{1}\left( kR\right) \left( e^{-k\left\vert \frac{L}{2}-z\right\vert
}-e^{-k\left\vert -\frac{L}{2}-z\right\vert }\right) \text{ ,}
\end{multline}%
where $J_{m}$ is a Bessel function of first kind and $m$ order.

Now it is possible to calculate the magnetostatic interaction energy between
two identical microwires using the magnetostatic field experienced by one of
the wires due to the other. The final results reads%
\begin{multline}
E_{int}=-\mu _{0}M_{0}^{2}\pi R^{2}L \\
\int_{0}^{\infty }\frac{dq}{q^{2}}e^{-q\left( 1+\frac{s}{L}\right)
}J_{0}\left( q\frac{d}{L}\right) J_{1}^{2}\left( q\frac{R}{L}\right)  \\
\left\{ 
\begin{array}{c}
\left( 1-e^{q}\right) ^{2}\qquad s\geq L \\ 
\left( 1-2e^{q}+e^{2q\frac{s}{L}}\right) \qquad s\leq L%
\end{array}%
\right. 
\end{multline}%
Equation (3) has been previously obtained for nanotubes. \cite{EAA+08} The
general expression for the interaction energy between wires with axial
magnetization, given by Eq. (3), can only be solved numerically. However,
wires that motivated this work satisfy $R/L=\alpha \ll 1$, in which case one
can use that $J_{1}\left( \alpha x\right) \approx \alpha x/2$. With this
approximation, Eq. (3) can be written in a very simple form as 
\begin{multline}
E_{int}=-\mu _{0}M_{0}^{2}\frac{\pi R^{4}}{4} \\
\left( \frac{1}{\sqrt{d^{2}+\left( L-s\right) ^{2}}}-\frac{2}{\sqrt{%
d^{2}+s^{2}}}+\frac{1}{\sqrt{d^{2}+\left( L+s\right) ^{2}}}\right) 
\end{multline}

Finally, the magnetostatic field can be written as a function of the
magnetostatic interaction between the magnetic microwires and is given by%
\cite{EAA+09}%
\begin{multline}
H_{int}=\frac{E_{int}}{\mu _{0}M_{0}V}=-\frac{M_{0}R^{2}}{4L} \\
\left( \frac{1}{\sqrt{d^{2}+\left( L-s\right) ^{2}}}-\frac{2}{\sqrt{%
d^{2}+s^{2}}}+\frac{1}{\sqrt{d^{2}+\left( L+s\right) ^{2}}}\right) 
\end{multline}

\section{Results}

Figure 3 shows the hysteresis loop for two microwires of radii of $R=6.5$ $%
\mu $m, length of $L=7$ mm, thickness of the glass coating of $t=3.3$ $\mu $%
m, and interaxial separation $d=45$ $\mu $m at room temperature. As
observed, the magnetization process takes place in two steps: a jump near $%
0.5$ Oe, that must be ascribed to the reversion of a microwire; and an
additional jump near $1.2$ Oe, corresponding to the reversion of the other
microwire. The flat region before the second step corresponds to the
magnetic field felt by one wire due to the other.

\begin{figure}[tph]
\begin{center}
\includegraphics[width=8cm]{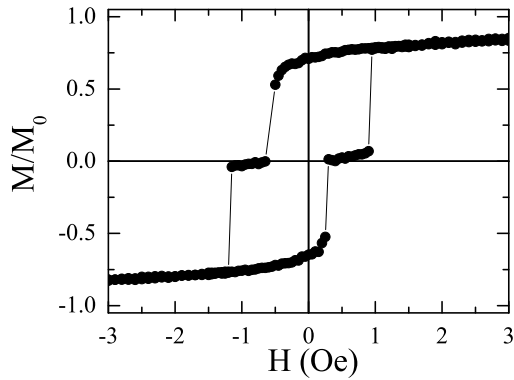}
\end{center}
\caption{Hysteresis loop for two parallel amorphous wires of Fe$_{77.5}$Si$%
_{7.5}$B$_{15}$.}
\end{figure}

\subsection{Dependence of the interaxial separation of the wires on the
magnetostatic field}

Theoretical and experimental magnetostatic results are combined in Fig. 4.
Experimental data for the magnetostatic field for different values of the
interaxial separation $d$ are depicted by gray dots and the theoretical
prediction is represented by the solid line. For the analytical calculations
we have used an effective radii of $R^{\ast }=4.7$ $\mu $m., which is
smaller than the real radius, in order to compensate the approximation of
consider a homogeneous magnetization in our model. \cite{PLK+07} It is clear
that there is a strong dependence of the magnetostatic field on $d$. Note
the good agreement between experimental datapoints and analytical results
for distances smaller than 80 $\mu $m. The coupling rapidly decreases by
increasing the separation between the wires, and become negligible for
interaxial distances bigger than 80 $\mu $m, at least within our
experimental sensitivity. It can be understood if we consider that the stray
field felt by one wire due to the other changes considerably if the wires
are not completely parallel to each other. Thus, when the separation between
the wires is big enough, small deviations may reduce the stray field
produced by the wires.

\begin{figure}[tph]
\begin{center}
\includegraphics[width=8cm]{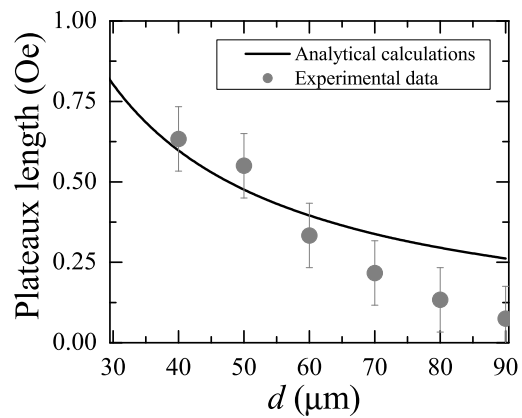}
\end{center}
\caption{Magnetostatic field as a function of the interaxial separation of
the microwires. The gray dots correspond to experimental data and the solid
line represents the values calculated analytically. Parameters: $R=6.5$ $%
\protect\mu $m, $L=7$ mm, $t=3.3$ $\protect\mu $m, $s=0$ $\protect\mu $m and 
$d$ ranging between $45$ $\protect\mu $m and $90$ $\protect\mu $m. For the
analytical calculations we have used an effective radii of $R^{\ast }=4.7$ $%
\protect\mu $m and $M_{0}=1.2\times 10^{6}$ A/m.}
\end{figure}

\subsection{Dependence of the horizontal separation of the wires on the
magnetostatic field}

It is interesting to analyze the behavior of the magnetostatic plateaux as
the interaxis distance, $d$, is kept fixed and the horizontal separation, $s$%
, is varied. Our results are combined in Fig. 5. Experimental data for the
magnetostatic field of the system are depicted by gray dots and the solid
line represents the analytical calculation. We observe a strong decrease of
the magnetic field as the horizontal separation between wires is increased.
Good agreement is obtained between the measured data and analytical
calculations.

\begin{figure}[tph]
\begin{center}
\includegraphics[width=8cm]{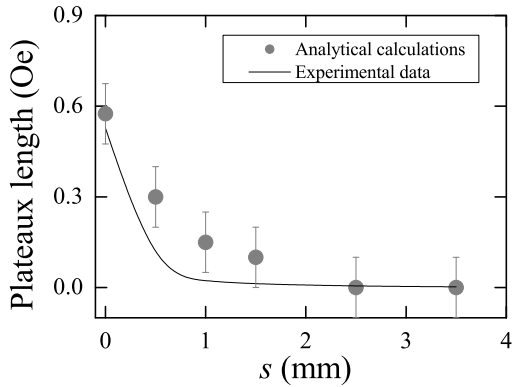}
\end{center}
\caption{Magnetostatic field as a function of the horizontal separation of
the microwires. The gray dots correspond to experimental data and the solid
line represents the values calculated analytically. Parameters: $R=6.5$ $%
\protect\mu $m, $L=7$ mm, $t=3.3$ $\protect\mu $m, $d=45$ $\protect\mu $m
and $s$ ranging between $0$ mm and $3.5$ mm. For the analytical calculations
we have used an effective radii of $R^{\ast }=4.7$ $\protect\mu $m and $%
M_{0}=1.2\times 10^{6}$ A/m.}
\end{figure}

\subsection{Interacting force}

A couple of isolated magnetic microwires can attract o repel each other upon
approach depending on their relative orientations. For the microwires
investigated the interacting force, $\mathbf{F}=-\mathbf{\nabla }E_{int}$,
has been estimated to be of the other of $0.017$ (for $d=90$ $\mu $m and $%
s=0 $ $\mu m$) to $0.0685$ dynes (for $d=45$ $\mu $m and $s=0$ $\mu m$) for
the case of interaxial separation in the range of parameters considered. For
the horizontal separation, the interacting force is in the range between $%
1.63\times 10^{-5}$ (for $d=45$ $\mu $m and $s=3.5$ mm) to $0.0685$ (for $%
d=45$ $\mu $m and $s=0$ mm) dynes.

\section{Discussion and Conclusion}

In summary, we have investigated the magnetostatic interaction between
magnetic microwires. Using a continuous model we have obtained a simple
expression to model the magnetostatic interaction in these particles. From
our calculations and measurements we can conclude that the magnetostatic
plateaux strongly depends on the geometry of the system. A couple of
isolated magnetic microwires can attract o repel each other upon approach
depending on their relative orientations. Based on microwires investigated
the interacting force, $\mathbf{F}=-\mathbf{\nabla }E_{int}$, has been
estimated to be of the other of $1.63\times 10^{-5}$ (for $d=45$ $\mu $m and 
$s=3.5$ mm) to $0.0685$ dynes (for $d=45$ $\mu $m and $s=0$ $\mu m$) in the
range of parameters considered. The perfomance of such experiment could
provide the basis for testing the different theoretical models. Our results
provide guidelines for the production of microstructures with tailored
magnetic properties.

\section{Acknowledgments}

The authors are grateful to R. Bernal for micrograph images and to D. Altbir
for useful discussions. This work was supported by Millennium Science
Nucleus \textit{Basic and Applied Magnetism} (project P06-022F), USAFOSR
(award N$^{o}$ FA9550-07-1-0040), and Fondecyt (Grants No. 11070010 and No.
1080164).

\end{document}